\documentclass[11pt,a4paper]{article}
\usepackage{jheppub}

\pdfoutput=1

\usepackage{mathtools}
\usepackage{amsmath, amssymb, amsthm, amsbsy, bbm}

\usepackage[percent]{overpic}
\usepackage{xcolor}
\usepackage[utf8]{inputenc}
\usepackage{enumitem}
\usepackage{hyperref}
\usepackage{graphicx}

\usepackage{subfig}
\usepackage{caption}
\usepackage[normalem]{ulem}
\usepackage{etoolbox}
\usepackage{cancel}
\usepackage{todonotes}

\newcommand{\ket}[1]{\left|#1\right\rangle}
\newcommand{\bra}[1]{\left\langle #1\right|}

\newcommand{\tr}{\mathrm{tr}}

\newcommand{\IR}{\mathrm{IR}}

 \newcommand{\braket}[2]{\langle #1 \vphantom{#2}|
    #2 \vphantom{#1} \rangle}

\theoremstyle{remark}

\theoremstyle{definition}

\newcommand{\calA}{\mathcal{A}}

\newcommand{\calC}{\mathcal{C}}

\newcommand{\calH}{\mathcal{H}}

\newcommand{\calL}{\mathcal{L}}

\newcommand{\calO}{\mathcal{O}}

\begin{document}

\title{Microstate Distinguishability, Quantum Complexity, and the Eigenstate Thermalization Hypothesis}

\author[1,2]{Ning Bao,}
\emailAdd{ningbao75@gmail.com}
\affiliation[1]{Computational Science Initiative, Brookhaven National
  Laboratory, Upton, New York, 11973}
\affiliation[2]{Center for Theoretical Physics and Department of Physics,
     University of California, Berkeley, CA 94720}
\author[3,4]{Jason Pollack,}
\emailAdd{jpollack@cs.utexas.edu}
\affiliation[3]{Department of Physics and Astronomy, University of British Columbia, Vancouver, BC V6T 1Z1, Canada}
\affiliation[4]{Quantum Information Center and Department of Computer Science, University of Texas at Austin, TX 78712}
\author[3]{David Wakeham}
\emailAdd{daw@phas.ubc.ca}
\author[2]{and Elizabeth Wildenhain}
\emailAdd{elizabeth\_wildenhain@berkeley.edu}

\abstract{
In this work, we use quantum complexity theory to quantify the difficulty of distinguishing eigenstates obeying the Eigenstate Thermalization Hypothesis (ETH). After identifying simple operators with an algebra of low-energy observables and tracing out the complementary high-energy Hilbert space, the ETH leads to an exponential suppression of trace distance between the 
coarse-grained eigenstates. Conversely, we show that an exponential hardness of distinguishing between states implies ETH-like matrix elements. The BBBV lower bound on the query complexity of Grover search then translates directly into a complexity-theoretic statement lower bounding the hardness of distinguishing these reduced states. 
}

\maketitle

\section{Introduction}

The Eigenstate Thermalization Hypothesis (ETH) \cite{Srednicki1994, Deutsch2018, DAlessio:2016rwt, Deutsch1991} is the conjecture that, under certain conditions, nearby energy eigenstates behave like states drawn from the microcanonical ensemble with respect to certain ``simple'' observables. Because random microcanonical fluctuations are suppressed by system size, we can interpret ETH as a conjecture about state indistinguishability: each state in an ETH ensemble is hard to distinguish from the ensemble average using simple observables.

The ETH is generally treated as a semi-empirical condition for state indistinguishability via measurement. As a rule of thumb, an ensemble with a density matrix that satisfies the ETH conditions (for some specification of simple observables) will behave like the microcanonical ensemble upon restriction to measurable observables, and its ensemble members will be indistinguishable via measurement. From the viewpoint of quantum information, however, these energy eigenstates are trivially distinguishable due to their orthogonality by the Holevo argument \cite{Holevo1973}. This might therefore seem to be a point of tension between the two approaches.

In this paper, we show how this tension may be resolved by making two main points. 
The first is that there is no conflict between the \textit{in-principle} perfect distinguishability of energy eigenstates and the \textit{in-practice} indistinguishability of eigenstates suggested by the ETH. As we will show, this difference can be simply understood using the information-theoretic language of quantum channels. Roughly, a macroscopic observer can be viewed as accessing the system only via a quantum channel which traces out fine-grained data about the system. We will demonstrate that for ETH ensembles this dramatically reduces the observed trace distance between states, realizing the operational constraints on the low-energy observer in terms of quantum information. Further, we partially invert this logic and deduce that exponential contraction of the trace distance between states implies ETH-like matrix elements. In our setup, a system where the ETH holds is thus roughly equivalent to a system where low-energy observers have difficulty telling things apart.

Our second main point is that ETH can be promoted from a semi-empirical belief to a formal complexity-theoretic statement about the difficulty of operationally distinguishing states after data restriction. We will show that the sharp lower bound on the complexity of Grover search given by BBBV \cite{BBBV} necessitates that distinguishing states in an ETH ensemble post-channel application takes exponentially many queries.  Effective indistinguishability can therefore be understood precisely in the language of quantum complexity theory.

An outline of our argument is as follows: 
\begin{enumerate}
    \item Restriction to an algebra of simple operators (representing coarse-grained or IR observables) is uniquely equivalent to a partial trace channel [Eq.~\eqref{eq:C_IR}].
    \item For an ETH ensemble, the partial trace channel exponentially suppresses the trace distance between ensemble members [Eq.~\eqref{eq:trace-sup}]. Conversely, exponential suppression 
    implies ETH-like matrix elements [Eq.~\eqref{eq:ETH-like}].
    \item Grover search distinguishes states by increasing their trace distance, thereby prying them apart. The procedure takes exponentially many queries to pry apart exponentially close states [Eq.~\eqref{eq:grover_iterations}] and is provably optimal for this task. The ETH is therefore itself lower-bounded (in our simple setup) to be exponentially hard by the lower complexity bound of distinguishing near-identical states implied by Grover search.
\end{enumerate}

In section \ref{sec:ETH-review}, we review the relevant aspects of ETH and its connection to distinguishability. In section \ref{sec:ETH-channel}, we describe our proposed coarse-graining quantum channel, discuss its connection to thermodynamics, and show that the channel exponentially suppresses the trace distance for an ensemble obeying the ETH. We also show the converse, namely that exponential suppression of trace distance implies ETH-like matrix elements.
In section \ref{sec:grover-exponential}, we review Grover search and its complexity bound, map it to the problem of state distinguishability, and show that Grover search requires exponentially many queries to distinguish exponentially suppressed states. This implies that restricting to coarse-grained observables in an ETH ensemble yields exponentially hard distinguishability, as well as the converse result that exponential hardness implies a simple form of the ETH. We finish with discussion and concluding comments in section \ref{sec:discussion}.

\section{A review of the Eigenstate Thermalization Hypothesis}
\label{sec:ETH-review}

\subsection{Statement of the Eigenstate Thermalization Hypothesis}

The Eigenstate Thermalization Hypothesis (ETH) was originally introduced as a conjecture about the conditions required for a quantum system to thermalize (i.e. exhibit expectation values that agree with the microcanonical ensemble) \cite{Srednicki1994}. The ETH states that the expectation values of observables for a quantum system with eigenstates $\ket{E_i}$ will evolve to those predicted by the microcanonical ensemble if the following two conditions on the observables are met: (i) the diagonal matrix elements  $\langle E_i|\hat{\calO}|E_i \rangle$ vary slowly with the state; and (ii) the off-diagonal matrix elements $\langle E_i|\hat{\calO}|E_j\rangle,~i\neq j$, are exponentially small in $N$, the number of degrees of freedom in the system \cite{Rigol2012}. 

In other words, the ETH is an ansatz for matrix elements of observables in the basis of the Hamiltonian's eigenstates. More formally, said ansatz is \cite{Rigol2012, Deutsch2018, DAlessio:2016rwt}:
\begin{gather}
    \calO_{ij} = \calO(\Bar{E})\delta_{ij}+e^{-S(\Bar{E})/2}f_\calO(\Bar{E},\omega)R_{ij}, \label{eq:ETH}
\end{gather}
where $\Bar{E}\equiv(E_i + E_j)/2$, $\omega \equiv E_i-E_j$, $S(E)$ is the thermodynamic entropy, and $\calO(\Bar{E})$ signifies the expectation value for the operator $\hat{\calO}$ in the microcanonical ensemble at energy $\Bar{E}$. Further, the ETH requires that $\calO(\Bar{E})$ and $f_\calO(\Bar{E},\omega)$ are smooth functions of $\Bar{E}$ and $\omega$, and that $R_{ij}$ behaves as a random variable with zero mean and unit variance (i.e. 
$\overline{|R_{ij}|^2}=1$). 

For an ensemble to satisfy the ETH, at least the vast majority of eigenstates must obey the above conditions. The ``weak ETH'' allows an exponentially small fraction of the eigenstates to violate the ETH, having significantly different expectation values from that of the microcanonical ensemble. On the other hand, the ``strong ETH'' states that $\calO_{ii}$ is very close to that of the microcanonical ensemble for \textit{all} the eigenstates. Because some models which do not thermalize (more precisely, some integrable models) satisfy the weak ETH, it is generally accepted that the strong ETH is required to characterize thermalization \cite{Deutsch2018}. For our purposes, however, it will not matter whether we use the strong or weak version of ETH.

\subsection{ETH and distinguishability}
The ETH condition on diagonal matrix elements is of the form \cite{Rigol2012, Deutsch2018, DAlessio:2016rwt}:
\begin{gather}
    \calO_{ii} = \calO(\Bar{E}) + R_{ii},
\end{gather}
where $R_{ii}$ is small (i.e., suppressed by the system size). Because this fluctuation of the expectation value from that of the microcanonical ensemble is very small for each eigenstate $\ket{E_i}$, each eigenstate is essentially indistinguishable from the ensemble average (and, by extension, from the other eigenstates).

Although ETH is not expected to hold for all operators, the general belief is that ETH applies to operators confined to a local region, which contain only a few degrees of freedom, and to low-point functions constructed from these operators, for most non-integrable systems \cite{Deutsch2018}. This claim is supported by semi-empirical evidence, such as numerical simulations of lattice systems \cite{DAlessio:2016rwt, Rigol2008, Santos2010, Santos2010b, Khatami2013, Rigol2009, Rigol2009b, Steinigeweg2013, Beugeling2014, Kim2014, Steinigeweg2014, Khodja2015, Beugeling2015, Biroli2010, Roux2010, Sorg2014, Neuenhahn2012,Khatami2012, Genway2012, Mondaini2016}, but the connection to low-energy restrictions on measurement and distinguishability has remained imprecise. In this work, we will provide a simple operator-algebraic interpretation of these low-energy operators, which formalizes and clarifies the expected loss of distinguishability.

\section{Coarse-graining the ETH}
\label{sec:ETH-channel}

In this section, we introduce the coarse-graining quantum channel for a low-energy ``algebraic" observer, and prove this channel is the unique coarse-graining compatible with our
assumptions. We also motivate our choice from the perspective of entropy maximization in thermodynamics. Finally, we discuss the effect of applying our partial trace channel to an ETH ensemble, showing that trace distance is exponentially suppressed, and derive a partial converse statement: that exponential contraction leads to an ETH-like expansion for matrix elements.\footnote{Our argument is similar in spirit to \cite{Dymarsky:2016ntg}, and for a single superselection sector, follows as a special case, where the subsystem $A$ is the low-energy factor.}

\subsection{Coarse-graining with quantum channels}
\label{sec:channel-IR}

A macroscopic observer interacting with a finite-dimensional
quantum system $\calH$ typically has access to some limited palette of coarse-grained observables, such as pressure and temperature in thermodynamics.
We will assume the simple observables form an (operator or von Neumann) \emph{algebra}, $\calA \subseteq \calL(\calH)$, containing the identity and closed under products and sums.
Thus, we have an ``algebraic" low-energy observer.

The \emph{Wedderburn decomposition} \cite{wedderburn1964lectures}
shows that such an algebra decomposes the full (assumed finite-dimensional) Hilbert space into irreducible
representations of $\calA$ as follows:
\begin{align}
    \calH & \simeq \left[\bigoplus_\alpha \calH_{1,\alpha} \otimes
  \calH_{2,\alpha}\right] \oplus \calH_0 \\ \quad M & = \left[\bigoplus_\alpha M_{1,\alpha} \otimes I_{2,\alpha}\right] \oplus 0\;, \label{eq:wedderburn}
\end{align}
for every $M \in \mathcal{A}$.
Here, $I_{2,\alpha}$ is the identity on $\calH_{2,\alpha}$, and
$M_{1,\alpha} \in \mathcal{L}(\calH_{1,\alpha})$.
The zero terms are present to handle the case in which none of the operators in $\calA$ are supported (i.e., act nontrivially) on some portion of the Hilbert space.
We also write $\calH_\alpha \equiv \calH_{1,\alpha}\otimes
\calH_{2,\alpha}$, with dimensions $d_{1,\alpha}, d_{2,\alpha}$, and decompose
arbitrary states as $\rho \equiv \bigoplus_\alpha p_\alpha \rho_\alpha \oplus p_0 \rho_0$.
The bracketed term in the Wedderburn decomposition was named a
\emph{generalized bipartition} by \cite{Kabernik2020}. The individual summands (labelled by $\alpha$) are analogous to superselection sectors
\cite{Kabernik2020} of commuting observables, whose associated projectors $\Pi_\alpha$, along with the zero projection, form a partition of unity, $I = \Pi_0 + \sum_\alpha \Pi_\alpha$.

\begin{figure}
    \centering
    \includegraphics[scale=0.52]{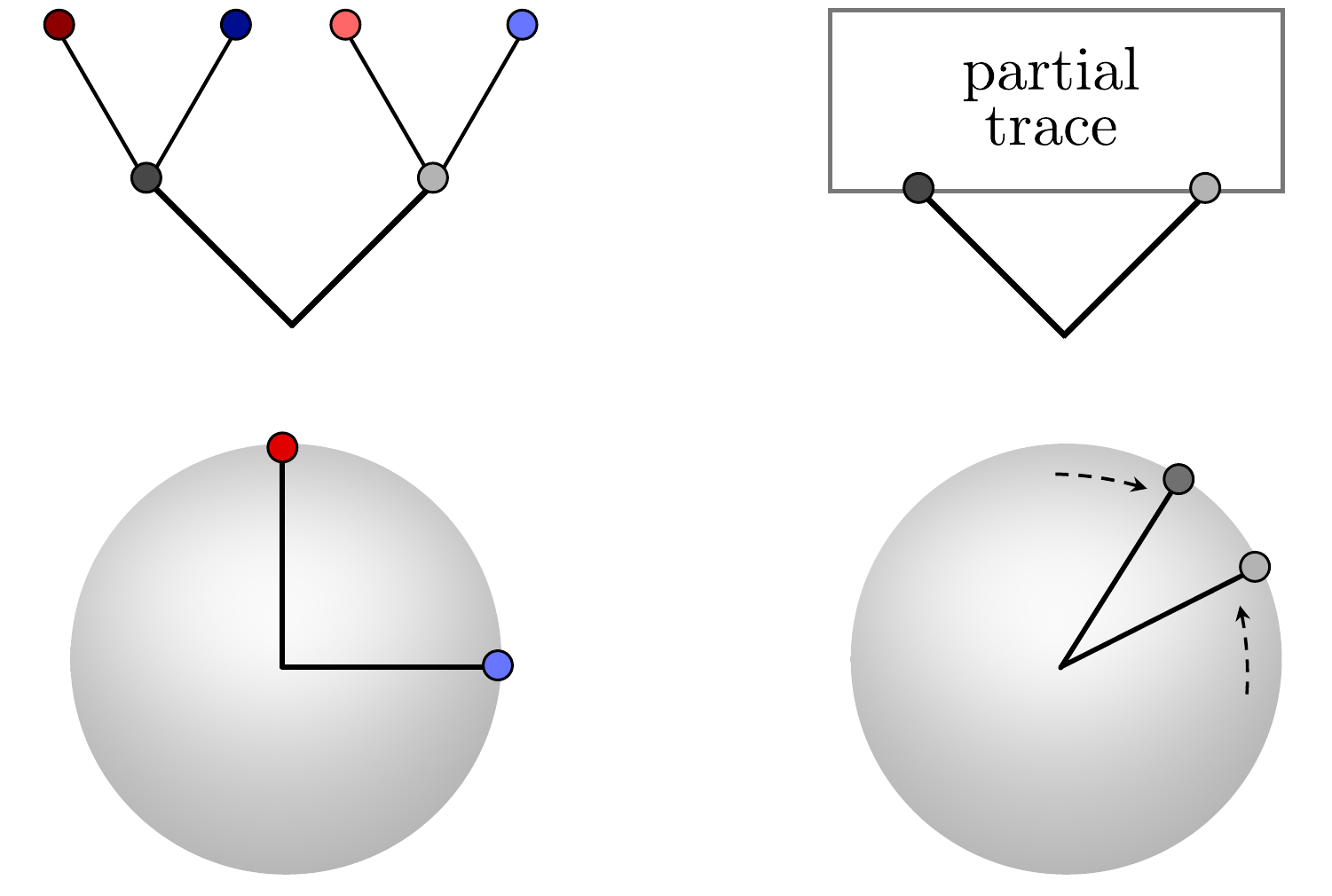}
    \caption{The coarse-graining in action, with shade representing the IR and hue the UV. Trace distance (section \ref{sec:ensemble}) between two states is also shown.
    \emph{Left.} Orthogonal states in the full tensor product. \emph{Right.} After hue is traced out, states are less distinguishable.}
    \label{fig:coarse}
\end{figure}

Here we will interpret each factor $H_{1,\alpha}$ as a macroscopic ``IR" Hilbert space, and each $H_{2,\alpha}$ as a space of fine-grained, macroscopically unobservable ``UV'' data.
Given this split, a natural quantum channel onto the IR is simply given by tracing out the UV Hilbert space $\calH_{2,\alpha}$ in each summand:
\begin{equation}
    \mathcal{C}_\IR: \rho \mapsto\overline{\rho} \equiv \bigoplus_\alpha p_\alpha
\tr_{2}[\rho_\alpha] \oplus p_0\rho_0\;.\label{eq:C_IR}
\end{equation}
It is easily confirmed that $\mathcal{C}_\IR$ is indeed a quantum channel: $\mathcal{C}_\IR$  is clearly linear and completely positive (it maps to a positive linear combination of densities), and we can check it is trace-preserving:
\[
\tr[\overline{\rho}] = p_0 + \sum_\alpha p_\alpha = 1 \;,
\]
since the coefficients $p_\alpha, p_0$ are normalized. Thus, $\mathcal{C}_\IR$  is a linear CPTP map and hence a quantum channel. 
We will focus for simplicity on states with nontrivial support in
a single sector $\alpha$, though our results easily generalize.
We will also largely ignore $\calH_0$, since no operator in $\calA$ has access to it. 
We picture the action of the partial trace on such a single sector in Fig.~\ref{fig:coarse}.

We note that our partial trace is equivalent\footnote{Up to subtleties due to identical particles that are irrelevant here \cite{Balachandran2013}.} to the restriction to the subalgebra $\calA$ \cite{Balachandran2013}.
Among quantum channels, the partial trace is therefore singled out as the \emph{minimal} way of discarding information about non-simple operators.
Coarse-graining could perform additional unitaries on the UV factors, but these are not constrained by the low-energy algebra and hence non-minimal.
We will give a thermodynamic justification for 
minimality 
in the next section.
One could consider something more complicated than a quantum channel, but the \emph{quantum Church-Turing thesis} \cite{qctt} conjectures that Nature is only as powerful as a quantum computer. A quantum channel is the only way such a computer has to discard everything but a subalgebra.

\subsection{Entropy maximization and tomographic completeness}
\label{sec:maximum-entropy}

We can equivalently view $\mathcal{C}_\IR$ 
as acting on the full Hilbert space.
Denote $\rho_{1,\alpha} \equiv \tr_{2,\alpha}[\Pi_\alpha \rho]$ and
$\rho_{2,\alpha} \equiv \tr_{1,\alpha}[\Pi_\alpha \rho]$.
Then $\mathcal{C}_\IR$ simply replaces each
$\rho_{2,\alpha}$ with the maximally mixed state $I_{2,\alpha}/d_{2,\alpha}$.
This relates our quantum channel interpretation of the ETH to the emergence of statistical mechanics from pure states, since the channel $\calC_\IR$ is closely related to \emph{Jaynes' maximum entropy principle} \cite{jaynes1}, which Katz elegantly formulated as ``the truth, and nothing but the truth" \cite{katz1967}. Suppose a quantum-mechanical system can be prepared in state $\rho$, and the observer measures some set
$\hat{\calO}\in A$.
Making measurements many times, they obtain a set of expectations (``the truth"):
\begin{equation}
\rho \mapsto  \{R_\calO(\rho) \equiv \tr[\hat{\calO}\rho]\}_{\hat{\calO}\in
  A}\;.\label{eq:1}
\end{equation}
Jaynes' principle states that the observer has most reason to believe
the system is in the \emph{maximum entropy state} (``nothing but the truth"):
\begin{equation}
\rho_{\text{MES}} = \mbox{argmax}_{R_\calO(\hat{\rho})=R_\calO(\rho)}
S(\hat{\rho})\;,\label{eq:2}
\end{equation}
where 
$S(\rho)$ is the \emph{von
  Neumann entropy}:
\begin{equation}
    S(\rho) \equiv -\tr[\rho \log \rho]\;.
\end{equation}
This can be maximized using Lagrange multipliers $\lambda_\calO$ and an auxiliary Gibbs ensemble \cite{katz1967}, yielding
\[
\rho_\text{MES} \equiv  \exp\left(\Omega - \sum_{\hat{\calO}\in A} \lambda_\calO \hat{\calO} \right)\;,
\]
for a normalization constant $\Omega$ obeying
\[
\Omega = -\log \tr \exp\left(-\sum_{\hat{\calO}\in A} \lambda_\calO \hat{\calO} \right)\;, \quad \frac{\partial\Omega}{\partial \lambda_\calO} = R_\calO\;.
\]
A derivation can be found in \cite{katz1967}.

In general, the map $\rho \mapsto \rho_{\text{MES}}$ 
is not linear.
But consider the set of observables $A$ consisting of all those of the form
\begin{equation}
    (\hat{\calO}\otimes I_{2,\alpha}) \oplus I_0 \bigoplus_{\beta\neq\alpha} I_\beta\;
\end{equation}
for some $\alpha$ and arbitrary $\hat{\calO} \in \calL{H_{1,\alpha}}$. We will abbreviate these operators as $\hat{\calO} \otimes I_{2,\alpha}$.
This is \emph{tomographically complete} for $\calH_{1,\alpha}$ in the sense that we uniquely recover $\rho_{1,\alpha}$ from the expectation values of the operators in $A_\alpha$.
In particular, we can use any orthonormal basis under the Hilbert-Schmidt inner product $\langle M_1, M_2\rangle \equiv \tr[M_1 M_2^\dagger]$ to directly reconstruct the density matrix.
To cap off Katz's phrase, tomographically complete sets capture ``the whole truth".
In this case, the Lagrange multiplier sum satisfies
\[
\sum_{A} \lambda_\calO (\hat{\calO}  \otimes I_{2,\alpha}) = \left(\sum_{A} \lambda_\calO \hat{\calO}\right) \otimes I_{2,\alpha}\;, \]
and the normalization factor
\[
\quad \Omega_\alpha = \Omega_{1,\alpha} - \log d_{2,\alpha} \;.
\]
Hence, the maximum entropy state is
\begin{align}
\Pi_\alpha \rho_\text{MES} & = \exp\left[\Omega_\alpha - \left(\sum_{A} \lambda_\calO \hat{\calO}\right) \otimes I_{2,\alpha}\right] \notag \\
& = \exp\left(\Omega_{1,\alpha} -\sum_{A} \lambda_\calO \hat{\calO}\right) \otimes \frac{I_{2,\alpha}}{d_{2,\alpha}} \notag \\
& =\rho_{1,\alpha} \otimes \frac{I_{2,\alpha}}{d_{2,\alpha}}\;,
\end{align}
using tomographic completeness on the first factor.

The first term captures ``the truth, the whole truth" (tomography on $\mathcal{H}_{1,\alpha}$) while the second factor captures ``nothing but the truth" (entropy maximization on $\mathcal{H}_{2,\alpha}$).
This result can be extended to the full sum over $\alpha$, since tomographic completeness with respect to an orthonormal basis allows us to reconstruct both the densities $\rho_\alpha$ on each $\alpha$ but also their coefficients $p_\alpha$, and entropy maximization proceeds on the second factor as before.
Thus, our partial trace is entropy-maximizing in the sense of Jaynes, and can therefore be interpreted as a simple thermodynamic coarse-graining.

\subsection{Contracting on ensembles}
\label{sec:ensemble}
Next we wish to demonstrate the effect of the coarse-graining channel on the distinguishability of states in an ETH ensemble.
A natural measure of the distinguishability of quantum states is the \emph{trace distance}, defined for densities $\rho, \sigma$ by
\begin{gather}
    D(\rho, \sigma)\equiv \frac{1}{2}\tr|\rho-\sigma|\;,
\end{gather}
where $|M| \equiv \sqrt{M^\dagger M}$.
In general, quantum channels $\mathcal{E}$ contract with respect to trace distance \cite{NandC}:
\begin{equation}
    D(\mathcal{E}(\rho), \mathcal{E}(\sigma)) \leq D(\rho, \sigma)\;.
\end{equation}
Suppose we select $\rho, \sigma$ from some ensemble of states $\mathbf{E}^{(\alpha)}$ in sector $\alpha$, and apply $\calC_\IR$.
Physically, we will interpret $\mathbf{E}^{(\alpha)}$ as the set of energy eigenstates spanning $\calH_\alpha$.

If the $\mathbf{E}^{(\alpha)}$ are exact eigenstates of the full Hamiltonian $H$, then $H$ must take diagonal form
\begin{equation}
    H = H_0 + \sum_\alpha H_\alpha\;, \label{eq:ham}
\end{equation}
where each $H_\alpha$ acts on $\calH_\alpha$.
This means that projectors $\Pi_\alpha$ commute with time evolution, and matrix elements vanish for eigenstates in different sectors.

We consider ensembles obeying the ETH, meaning that ``simple" operators $\hat{\calO} \otimes I_{2,\alpha}$, for $|E_i\rangle, |E_j\rangle \in \mathbf{E}^{(\alpha)}$, have matrix elements of the form
\begin{equation}
\langle E_i |\hat{\calO} \otimes I_{2,\alpha} |E_j \rangle = \calO(\bar{E})\delta_{ij} + f_\calO^{(\alpha)}e^{-S/2} R_{ij}\;.
\label{eq:general-O}
\end{equation}
Again, we will focus on a single sector $\alpha$, with $S = \log d = \log d_{1,\alpha} d_{2,\alpha}$.
We will take the size of the low-energy Hilbert space $d_{1,\alpha}$ to be small and fixed, and $d_{2,\alpha}$ to be large, so that asymptotic growth with respect to system size implies fixed $d_{1,\alpha}$ and increasing $d_{2,\alpha}$, so that $d_{2,\alpha} = \Omega(e^S)$ and $d_{1,\alpha} = \Omega(1)$.

As a warm-up, suppose the energy eigenstates $\mathbf{E}^{(\alpha)}$ are Haar-random, i.e.
obtained from a reference state
$\ket{\psi^{(\alpha)}}$ by applying $k$ independent unitaries $U_i \in
\mathrm{U}(\calH_\alpha)$ chosen with Haar measure.
Page's theorem \cite{Page_1993} states that Haar-random states
$\rho_{1,\alpha}(U) \equiv \calC_\IR (U \ket{\psi^{(\alpha)}})$ are
close to maximally mixed: 
\begin{align}
\int \mathrm{d}U\, D\left(\rho_{1,\alpha}(U),
  \frac{I_{1,\alpha}}{d_{1,\alpha}}\right) & \leq 
\frac{1}{2}\sqrt{\frac{d_{1,\alpha}^2 - 1}{d_{1,\alpha}d_{2,\alpha} +
  1}} \notag \\
& \leq \frac{1}{2}\sqrt{\frac{d_{1,\alpha}}{d_{2,\alpha}}} \;.
\end{align}
From the triangle inequality for $D$, we find a bound on the
Haar-averaged trace distance for $\rho, \sigma \in
\mathbf{E}^{(\alpha)}$, after applying $\calC_\IR$:
\begin{equation}
\overline{D\left(\calC_\IR(\rho), \calC_\IR (\sigma)\right)} \leq \frac{1}{2} \sqrt{\frac{d_{1,\alpha}}{d_{2,\alpha}}}\;.
\end{equation}
From our assumptions $d_{2,\alpha} = \Omega(e^{S})$ and $d_{1,\alpha} = \Omega(1)$ in system size, we see that typical random states have their trace distance suppressed on the order of $e^{-S/2}$.

Subject to our assumptions about the dimension of the IR and UV factors, we will find a similar suppression from the perspective of the low-energy algebra in a moment.
However, we note that for a generic set of IR observables, these assumptions are tantamount to taking an exponentially thin energy shell.\footnote{We thank Anatoly Dymarsky for discussion of this point.}
This is not, in general, a realistic assumption for thermal systems, and more work is required to understand if (and whether) the algebraic decomposition can be used to make the sorts of constraints we are considering here.
However, subject to these assumptions, the smoothness of the microcanonical average $\calO(E)$ implies that differences in the leading-order variation of the diagonal terms in (\ref{eq:general-O}),
\[
\calO(E_i) - \calO(E_j) \approx \calO'(E)\, \omega\;,
\]
is also exponentially suppressed.
Thus, fluctuations of the microcanonical average contribute at subleading order in (\ref{eq:general-O}), and we can restrict to matrix elements of the form
\begin{equation}
\langle E_i |\hat{\calO} \otimes I_{2,\alpha} |E_j \rangle = \overline{\calO}^{(\alpha)}\delta_{ij} + f_\calO^{(\alpha)}e^{-S/2} R_{ij}\;.
\label{eq:flat-O}
\end{equation}
This is analogous to the exponentially thin ETH ensemble considered in \cite{Dymarsky:2016ntg}. We call such superselection sectors \emph{flat}.
We will comment briefly on the more general case below.

By restricting to observables of the form $\hat{\calO}\otimes I_{2,\alpha}$ and considering expectations, we implicitly apply our channel.
Defining $\overline{\rho} \equiv \calC_\IR(\rho)$, and $\rho_i \equiv |E_i\rangle\langle E_i|$, note that
\[
\langle E_i | \mathcal{O}\otimes I_{2,\alpha}|E_i\rangle = \tr[\rho_i (\hat{\calO}\otimes I_{2,\alpha})] = \tr[\overline{\rho}_i \hat{\calO}]\;,
\]
since expectations on a tensor factor are given by expectations with respect to the reduced density. This is easily verified using the Schmidt decomposition.
Thus, we have
\begin{align}
    \tr[(\overline{\rho}_i-\overline{\rho}_j) \hat{\calO}] & = \langle E_i | \hat{\calO}\otimes I_{2,\alpha}|E_i\rangle - \langle E_j | \hat{\calO}\otimes I_{2,\alpha}|E_j\rangle \notag \\
    & = f^{(\alpha)}_\calO e^{-S/2} (R_{ii}-R_{jj})\;. \label{eq:squish}
\end{align}
To relate this to trace distance, we need to recall its variational form \cite{NandC},
\begin{align}
  D(\rho, \sigma) & = \frac{1}{2}\max_{-I \leq \hat{\calO} \leq I} \tr [(\rho -
                    \sigma)\hat{\calO}] \label{eq:varO}\;, %
\end{align}
where $A \leq B$ means $B - A$ is positive semidefinite, or equivalently, the eigenvalues of $\hat{\calO}$ lie between $-1$ and $1$.
If we unit normalize the operators $\hat{\calO} \in \calA$, and assume the variance $f_\calO^{(\alpha)} = \Omega(1)$ in the system size,
then the trace distance is controlled by the maximum of random diagonal matrix elements in (\ref{eq:squish}).
(Note that the maximum is attained by an observable of the form $\hat{\calO}\otimes I_{2,\alpha}$ since the operators are of this form by (\ref{eq:wedderburn}).)

Although $R_{ij}$ has mean zero and unit variance by assumption, we are making $d = e^{S}$ independent draws, and the maximum will depend on $S$.
We can define the expected maximum value $x_d$ as the point where the tail of the cumulative distribution function (cdf) equals $1/d$:
\[
1 - F(x_d) = \frac{1}{2}\left[1-\frac{1}{\sqrt{2}}\mbox{erf}(x_d)\right] = \frac{1}{d}\;,
\]
where $F(x)$ is the cdf of $R_{ii}$, presumed Gaussian.
Taking $d$ and hence $x_d$ large, the standard large argument asymptotics for $\mbox{erf}(x)$ give 
\[
    x_d \sim \sqrt{\log d} = \sqrt{S}\;.
\]
There are subleading corrections we can ignore.
In the large sample limit, the Fisher-Tippett theorem \cite{reiss2007} shows this estimate is asymptotically sharp, and independent of the nature of the identically distributed zero mean, unit variance draws.\footnote{This is analogous to the central limit theorem, so it is sometimes called the \emph{max central limit theorem}. There are various technical niceness conditions on the distributions, but we will not belabor them here.} 
Combining with (\ref{eq:squish}), we find
\begin{equation}
    D(\overline{\rho}_i,\overline{\rho}_j) = \Omega(\sqrt{S}e^{-S/2})\;. \label{eq:trace-sup}
\end{equation}
One can set a pair of constants $k$ and $k'$ in the exponent such that $e^{-kS}\leq \sqrt{S}e^{-S/2} \leq e^{-k'S}$ as $S$ becomes asymptotically large.
Thus, passing the flat ETH ensemble through our quantum channel results in exponential contraction.

For the more general, i.e non-flat, condition (\ref{eq:general-O}), a similar calculation gives the trace distance for $\overline{\rho}_i, \overline{\rho}_j$ in terms of the size of the energy window $\Delta E$ and the system size:
\[
D(\overline{\rho}_i, \overline{\rho}_j) = \Omega(\omega) + \Omega(e^{-S/2})\;,
\]
Thus, in general, the suppression of trace distance will depend 
width of the energy windows corresponding to a superselection sector.
For an $O(1)$ energy window, we generically expect \emph{polynomial} suppression of trace distance
\[
D(\overline{\rho}_i, \overline{\rho}_\text{micro}) = O \left(\frac{1}{p(S)}\right)\;,
\]
for some polynomial in $S$ \cite{Dymarsky:2016ntg}.

\subsection{A partial converse}

We can ask whether the converse holds, i.e. that exponential suppression of trace distance 
implies an ETH-like form for the matrix elements.
It is clear from (\ref{eq:varO}) and (\ref{eq:trace-sup}) that diagonal matrix elements for energy eigenstates can only \emph{differ} by terms suppressed by $e^{-S/2}$.
This means we have the flat diagonal expectations, as per (\ref{eq:flat-O}).
To fix the leading order behaviour, we simply note that these diagonal elements are close to each other, and hence the ensemble average $\overline{\calO}^{(\alpha)}$.
More carefully, we can define a microcanonical density
\[
\rho_{\text{micro}} = e^{-S} \sum_i |E_i\rangle\langle E_i| = e^{-S} \sum_i \rho_i \;.
\]
Then, by linearity of the channel, $\overline{\rho}_\text{micro} = e^{-S}\sum_j \overline{\rho}_j$ and hence
\begin{align}
D(\overline{\rho}_i, \overline{\rho}_\text{micro})  & = \frac{1}{2}\sup_{\hat{\calO}} \tr[(\overline{\rho}_i - \overline{\rho}_\text{micro})\hat{\calO}] \notag \\ & = \frac{1}{2}\sup_{\hat{\calO}} \sum_j e^{-S} \tr[(\overline{\rho}_i -\overline{\rho}_j)\hat{\calO}] = \Omega(e^{-S/2})\;.\notag
\end{align}
Thus, for any eigenstate $i$, $\langle \hat{\calO}\rangle_i \approx \overline{\calO}$ up to corrections of order $e^{-S/2}$.

We still need to constrain off-diagonal elements.
Let us first consider the implications of the ETH assumption. Define the ``rotated" states $|E_{i\pm j}\rangle \equiv (|E_i\rangle \pm |E_j\rangle)/\sqrt{2}$, with densities
\begin{equation}
    \rho^\pm_{ij} \equiv |E_{i\pm j}\rangle \langle E_{i\pm j}|\;.
\end{equation}
Then the difference of these densities gives the off-diagonal elements:
\begin{equation}
\rho^+_{ij} - \rho^-_{ij} = |i\rangle\langle j|+ |j\rangle\langle i|\;.
\end{equation}
The ETH assumption implies
\begin{align}
\tr[(\rho^+_{ij} - \rho^-_{ij})(\hat{\calO} \otimes I_{2,\alpha})] = \tr[(\overline{\rho}^+_{ij} - \overline{\rho}^-_{ij})\hat{\calO}]\notag
 = f_\calO^{(\alpha)}e^{-S/2} (R_{ij} + R_{ji})\;,
\end{align}
which by similar arguments is $\Omega(e^{-S/2})$.
Going in the reverse direction, we learn that if the trace distance between $\rho^\pm_{ij}$ is exponentially contracted, then
\begin{equation}
    \langle i | \hat{\calO} | j \rangle + \langle i | \hat{\calO}^\dagger | j \rangle = O(e^{-S/2})\;.
\end{equation}
For Hermitian $\hat{\calO}$, this is precisely the scaling we expect for off-diagonal ETH elements.

Thus, combining the trace distance constraints on the reduced densities for $|E_i\rangle, |E_j\rangle \in \mathbf{E}^{(\alpha)}$, we have the ETH-like matrix elements for Hermitian operators\footnote{This resembles the Knill-Laflamme-like \cite{Knill_2000} condition for approximate quantum error correcting codes (AQECC) given in \cite{Brand_o_2019}. However, in our case, the natural code subspace $\calH_\alpha$ also controls the ETH suppression (rather than the full Hilbert space), so we do not obtain a good AQECC.}
\begin{equation}
    \langle i | \hat{\calO}| j \rangle = \overline{\calO}^{(\alpha)} \delta_{ij} + e^{-S/2}A_{ij}\;, \label{eq:ETH-like}
\end{equation}
for $A_{ij} = O(1)$ in system size.
For a general, i.e. non-flat superselection sector, leading-order variations in the micronical average $\calO(E)$ prevent this clear-cut identification of terms in an $e^{-S}$ expansion.

\section{
Distinguishing exponentially close states
}
\label{sec:grover-exponential}
The results of the previous section show that application of our coarse-graining channel to an ETH ensemble yields exponentially-suppressed trace distance among the members of the resulting ensemble. We were further able to establish a partial converse result: a coarse-graining map that yields exponential suppression implies ETH-like matrix elements for simple operators.

In this section, we move from information-theoretic to complexity-theoretic considerations. We connect the query complexity lower bound of Grover search to the hardness of distinguishing ETH ensemble states after coarse-graining. The Grover search algorithm \cite{Grover1996} is a quantum search algorithm for a marked item in an unstructured data set. As will be discussed in the review below, it is quadratically faster than the fastest known classical algorithm (which runs in $O(N)$ time), running in $O(\sqrt{N})$ time. 

While this modest but important speedup would usually not be considered significant in quantum algorithms, its importance lies in the fact that it is provably tight: both the query complexity---the number of queries an algorithm must make before reaching an answer---and the gate complexity---the number of required gates---of Grover search are the most efficient theoretically possible for the unstructured search problem, even up to the leading pre-factor \cite{BBBV}. As such, if something requires violating the query complexity or gate complexity lower bound for Grover search, it is often tantamount to a significant modification of quantum mechanics, as discussed for example in \cite{Bao2016}.

In particular, we note that because distinguishability of the pure states is easier than mixed state distinguishability for mixed states, we restrict our attention in searching for a lower bound on ETH distinguishability by focusing on hypothetical pure states that are outputs of the coarse graining quantum channel. This is not to ascribe any physical significance to the Grover search algorithm used or the existence of the reflection operator therein, but rather to use this method to provide a conservative but still substantial lower bound to the complexity of ETH distinguishability.

\subsection{Review of Grover's search algorithm}
We begin with a review of Grover search in the context of the simple search problem it was originally designed to solve \cite{Grover1996, NandC}. Suppose one aims to find a particular element in an unsorted set of size $N$. Given that the set has no structure, the most efficient classical method is to cycle through all the elements in the set one-by-one. Therefore, the classical solution to this search problem runs in $O(N)$ time.

Grover search is a quantum algorithm that solves the same search problem with only $O(\sqrt{N})$ operations. An outline of the algorithm is as follows \cite{NandC}:
\begin{enumerate}[label=\alph*.]
    \item Start with the state $\ket{0}^{\otimes n}$.
    \item Apply the Hadamard transform, which produces the superposition \newline state $\ket{\psi} = \frac{1}{N^{1/2}}\sum_{x=0}^{N-1}\ket{x}$.
    \item Repeatedly apply Grover iteration.
\end{enumerate}
Grover iteration consists of the following steps:
\begin{enumerate}
    \item Apply the ``oracle" operator $O$, which marks the desired item by shifting its phase. The oracle takes the state $\ket{x}$ to $(-1)^{f(x)}\ket{x}$, where $f(x)=1$ if $x$ is a solution to the problem and $f(x) = 0$ otherwise.
    \item Apply the Hadamard transform.
    \item Perform a phase shift of $-1$ on all states except $\ket{0}$. One can achieve this by acting on the state with the operator $2\ket{0}\bra{0}-I$.
    \item Apply the Hadamard transform again.
\end{enumerate}
In summary, Grover iteration is an application of the operator 
\begin{equation}
    G = \left(2\ket{\psi}\bra{\psi}-I\right)O.
\end{equation}

\begin{figure}
    \centering
    \includegraphics[width=.3\textwidth]{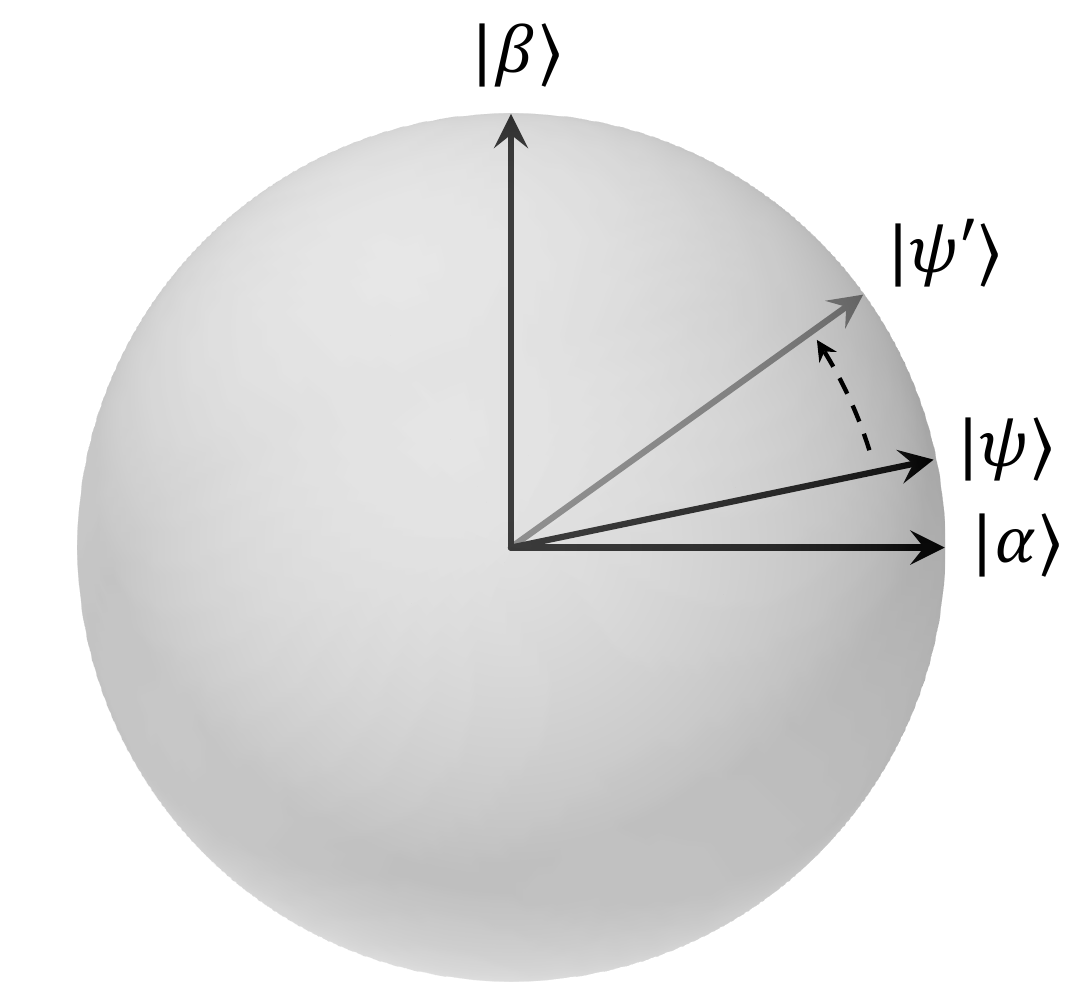}
    \caption{Geometric visualization of Grover iteration. States are represented as vectors on the Bloch sphere. Each application of Grover iteration rotates the initial state $\ket{\psi}$ toward the normalized sum of solutions to the search problem, $\ket{\beta}$.}
    \label{fig:2}
\end{figure}

If one visualizes quantum states as vectors on the Bloch sphere, Grover iteration has an intuitive geometric visualization (see Fig.~\ref{fig:2}) \cite{NandC}. We can re-express the starting quantum state $\ket{\psi}$ in terms of (i) a state $\ket{\beta}$ that is the normalized sum of solutions to the search problem and (ii) a state $\ket{\alpha}$ that is the normalized sum of states that are \textit{not} solutions to the search problem. Defining an angle $\theta$ in terms of the total number of states ($N$), and the number of solutions to the search problem ($M$) via $\cos(\theta/2) = \sqrt{(N-M)/N}$, the initial state is:
\begin{gather}
    \ket{\psi}= \cos\frac{3\theta}{2}\ket{\alpha} + \sin\frac{3\theta}{2}\ket{\beta}.
\end{gather}
Each application of the Grover iteration operator ($G$) effects a rotation in the space spanned by $\ket{\alpha}$ and $\ket{\beta}$ by the angle $\theta$. After $k$ Grover iterations, the rotated state is
\begin{gather}
    \ket{\psi'}=\cos\left(\frac{2k+1}{2}\theta\right)\ket{\alpha} + \sin\left(\frac{2k+1}{2}\theta\right)\ket{\beta}.
\end{gather}
Thus, repeated Grover iterations rotate the state closer and closer to $\ket{\beta}$, the sum of solutions to the search problem.

Once the component parallel to $\ket{\beta}$ is greater than the component parallel to $\ket{\alpha}$, a measurement is more likely to produce a solution to the search problem than not. Expressing the nearest integer to some real number $x$ as the function $\mathrm{CI}(x)$, the number of Grover iterations required to achieve this is
\begin{gather}
    R = \mathrm{CI}\left(\frac{\arccos\sqrt{M/N}}{\theta}\right)\leq \Big\lceil \frac{\pi}{4}\sqrt{N/M} \Big\rceil.
\end{gather}
Grover search therefore requires $O(\sqrt{N/M})$ iterations (or $O(\sqrt{N})$ iterations if there is only one solution to the search problem), making it more efficient than the classical algorithm \cite{Grover1996, NandC}.

What makes Grover search of particular interest is that the complexity bound of $O(\sqrt{N})$ is provably optimal; no other search algorithm can complete the task in fewer operations, nor with fewer queries \cite{BBBV, NandC}. A sketch of the proof is as follows. Suppose we have some quantum algorithm that applies the oracle for a given search solution $x$ and some set of unitary operations $U_i$ such that, after $k$ applications of the oracle, it produces the state
\begin{gather}
    \ket{\psi_k^x}\equiv U_k O_x U_{k-1}O_x...U_1 O_x \ket{\psi}.
\end{gather}
One can prove that Grover search is optimal by examining the magnitude of the effect of the oracle, or, more precisely, the deviation from the state that would have evolved in the absence of the oracle. Defining the state evolved without the oracle as
\begin{gather}
    \ket{\psi_k}\equiv U_k U_{k-1}...U_1\ket{\psi},
\end{gather}
the deviation after $k$ steps is defined as
\begin{gather}
    D_k \equiv \sum_x \big | \big | \psi_k^x - \psi_k \big | \big |^2.
\end{gather}
If this deviation is small, then the component parallel to $\ket{\beta}$ is not yet larger than the component parallel to $\ket{\alpha}$. Therefore, a small deviation implies that a solution to the search problem is not yet identifiable. The proof of optimality requires showing that (i) the deviation after $k$ steps ($D_k$) obeys $D_k \leq 4k^2$ and (ii) the probability of identifying the search solution is greater than 1/2 only if $D_k$ is $\Omega(N)$. Taken together, (i) and (ii) imply that the requisite number of oracle calls obeys $k\geq \sqrt{cN/4}$. Therefore, solving the search problem requires calling the oracle $\Omega(\sqrt{N})$ times. As this is the complexity lower bound of Grover search, Grover search is an optimal solution to the search problem \cite{BBBV, NandC}.

\subsection{Grover search for distinguishing states with exponentially suppressed trace distance}
\begin{figure}
    \centering
    \includegraphics[width=.3\textwidth]{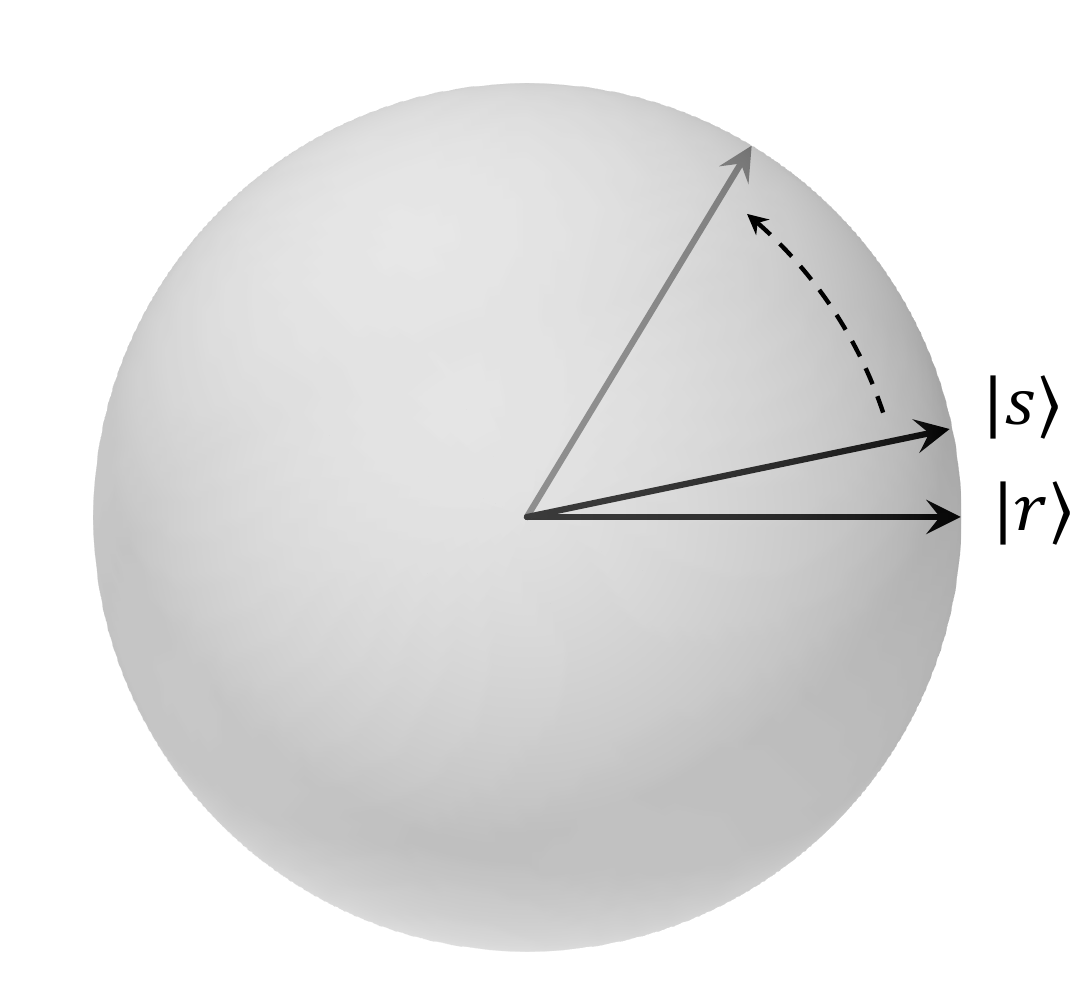}
    \caption{Geometric visualization of distinguishing states $\ket{s}$ and $\ket{r}$ via Grover search. States are represented as vectors on the Bloch sphere. Grover iteration rotates $\ket{s}$ away from $\ket{r}$.}
    \label{fig:3}
\end{figure}

Although the above review of the Grover algorithm is framed in terms of a search problem, the method translates to the problem of distinguishing two quantum states. In the latter case, the two states to be distinguished, $\ket{r}$ and $\ket{s}$,  play the roles of $\ket{\alpha}$ and $\ket{\psi}$ respectively. The function of Grover iteration is to rotate state $\ket{s}$ away from $\ket{r}$ until a measurement produces a state that is not $\ket{r}$ with probability greater than 1/2. See Fig.~\ref{fig:3} for a visualization on the Bloch sphere.

The number of Grover iterations required to pry apart the states depends on how similar the states were originally. As mentioned previously, a measure of the similarity between (or distinguishability of) two quantum states is the trace distance, defined for two states with density matrices $\rho$ and $\sigma$ as
\begin{gather}
    D(\rho, \sigma)\equiv \frac{1}{2}\tr|\rho-\sigma|,
\end{gather}
where $|M| \equiv \sqrt{M^\dagger M}$  \cite{NandC}. If the two states are single qubits, one can express the trace distance in terms of their Bloch vectors $\overrightarrow{r}$ and $\overrightarrow{s}$ as:
\begin{gather}
    D(\rho, \sigma) = \frac{|\overrightarrow{r}-\overrightarrow{s}|}{2}.
\end{gather}
Thus the trace distance for qubits is the Euclidean distance between their Bloch vectors up to a multiplicative factor. In other words, smaller trace distance between two states means they are closer together on the Bloch sphere.

Our aim is to find a lower bound on the hardness of distinguishing states after application of the channel defined in section \ref{sec:channel-IR}. To do so, we must compute roughly how many Grover iterations are required to distinguish $\ket{s}$ from $\ket{r}$ given that their trace distance is exponentially suppressed. To relate the initial trace distance between the vectors to the number of Grover iterations required to pry them apart, we note that the initial trace distance $D$ is related to the initial angle $\theta_{rs}$ between the Bloch vectors $\overrightarrow{r}$ and $\overrightarrow{s}$ via
\begin{gather}
    D^2 = \frac{1}{4}|\overrightarrow{r}-\overrightarrow{s}|^2 = \frac{1}{2}(1-\cos\theta_{rs})\label{eq:grover_iterations}
\end{gather}
assuming $\overrightarrow{r}$ and $\overrightarrow{s}$ are normalized. We are interested in the case where the states $\ket{r}$ and $\ket{s}$ are very similar, so we can expand $\cos\theta_{rs}$ to see that 
\begin{align*}
    1-\frac{\theta_{rs}}{2} &\approx 1-2D^2\\
      &\Longrightarrow \quad \theta_{rs} =  2D.
\end{align*}
As defined in the previous section, the rotation angle of Grover iteration is related to the initial angle between $\overrightarrow{r}$ and $\overrightarrow{s}$ by $\theta = 2\theta_{rs}$. Successfully distinguishing the states occurs when the angle between $\overrightarrow{r}$ and the post-rotation vector $\overrightarrow{s}'$ obeys $\theta_{rs'} = \frac{\theta}{2}\geq \frac{\pi}{4}$. Assuming that the initial angular separation $\theta_{rs}$ is small, this will require rotating through approximately $\frac{\pi}{4}$ radians. Therefore, we must apply approximately $\frac{\pi}{16D}$ Grover iterations to distinguish the states. Assuming $D$ is exponentially small, this corresponds to exponentially many Grover iterations.\footnote{If this suppression is only inverse polynomial, this requires correspondingly polynomially many Grover iterations. For states in quantum field theories of very large numbers of qubits, this is still in practice a very long run time.} Therefore, because Grover search is optimal, distinguishing two states with exponentially suppressed trace distance requires exponentially many queries. 

Let us now consider the states that are separated from their ensemble average by an exponentially suppressed trace distance, after the partial trace channel has been implemented.
Although the above discussion focused on single-qubit states, it is a straightforward extension of this result that mixed states on more than one qubit (as opposed to the single qubit described by the Bloch sphere) will similarly require an exponential number of queries to distinguish between them, as the amplitude amplification aspect of Grover search functions in an identical manner. The BBBV result \cite{BBBV} still applies to the mixed-state amplification problem \cite{Biham_2002}, as the difficulty of distinguishing the mixed states which are the generic output of the coarse-graining channel is lower-bounded by the difficulty of distinguishing pure states in the IR Hilbert space by the data processing inequality. Recall that the ETH states that upon restricting to low point functions (e.g. applying the partial trace quantum channel) it becomes ``hard" to distinguish a state in an ETH-obeying ensemble from the ensemble average. Relating this difficulty to the query complexity of Grover search is a straightforward way of quantifying this ``hardness" in more precise language. As such, this is a complexity-theoretic way of seeing why states in ETH ensembles are difficult to distinguish from each other, especially once the additional overhead from implementing the data processing inequality is considered.\footnote{It is worth noting here that we do not consider the complexity of implementing the quantum channel, this result should technically only be considered an argument that the difficulty of distinguishing such states is \emph{at least} exponentially hard.}

Elegantly, there is also room for non-ETH states in this picture; they are simply the states that, upon the partial trace operation, are not exponentially close in trace distance from some canonical state that one is attempting to distinguish it from.
\begin{figure*}
    \centering
    \includegraphics[width=\textwidth]{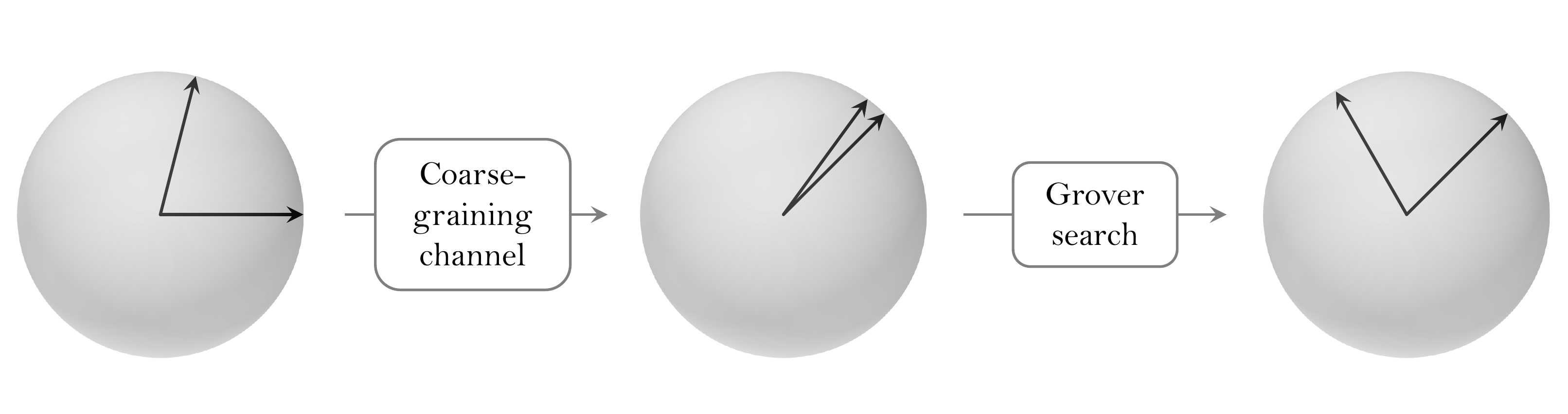}
    \caption{Classifying the hardness of ETH with quantum channels. A partial trace channel coarse-grains UV data and suppresses trace distance for some ensembles. Grover search pries apart states in the ensemble and makes them distinguishable.}
    \label{fig:4}
\end{figure*}

Conversely, it is immediate from the tightness of the BBBV lower bound that if density matrices take exponentially many queries to distinguish using Grover search, they must be exponentially close in trace distance. In particular, if this holds both for the reduced densities of eigenstate $\overline{\rho}_i = \calC_\IR(|E_i\rangle\langle E_i|)$, $|E_i\rangle \in \mathbf{E}^{(\alpha)}$, and ``rotated" reduced densities 
$\overline{\rho}^\pm_{ij} = \calC_\IR(|E_{i\pm j}\rangle\langle E_{i\pm j}|)$, then the results of section \ref{sec:ensemble} immediately give the ETH-like expansion of matrix elements (\ref{eq:ETH-like}).

\section{Discussion}
\label{sec:discussion}
\subsection{Summary}
Let us briefly summarize our arguments.
First, we showed that the orthogonality (and hence perfect distiguishability \`{a} la Holevo \cite{Holevo1973}) of eigenstates is compatible with near indistinguishability when passed through a quantum channel.
In our model, the quantum channel is a generalized partial trace, and the algebra of observables on the remaining Hilbert space factors the set of simple operators for the purposes of ETH.
This channel has a simple thermodynamic interpretation in the spirit of entropy maximization and Jaynes' principle.
Furthermore, for an ensemble of eigenstates obeying the ETH, exponential suppression of energy differences in a coarse-grained window results in exponential suppression of trace distance.

Although closeness in trace distance already suggests hardness of distinction from a quantum information-theoretic viewpoint, we further lower bound this hardness via a combination of the data-processing inequality and a complexity-theoretic perspective using Grover search. The BBBV lower bound on search algorithms \cite{BBBV} shows the optimality of Grover search, and hence the task of telling apart our exponentially close ETH states will require at least an exponential number of queries. See Fig.~\ref{fig:4} for a visual summary.

Finally, we were able to partially reverse the logic of these steps. If states require an exponential number of queries to distinguish, it follows they are exponentially close in trace distance.
If our coarse-graining channel exponentially suppresses the trace distance between reduced eigenstates, then they exhibit ETH-like matrix elements for simple operators.
Loosely speaking, hardness of distinction is equivalent to the ETH in a thin energy band.

\subsection{RG flow and time evolution}

In field-theoretic settings one often expresses ignorance of the behavior of the UV and any unknown massive fields that may live there by ``integrating out'' or, more formally, with arguments using renormalization group (RG) flow. We briefly comment on the differences between this textbook approach and the coarse-graining methods we employ here. (To avoid complication, we have in mind a field theory that has been regularized, e.g.\ on a lattice, so that all Hilbert spaces under discussion are finite-dimensional.) When a Hilbert space describes a field theory, we can describe the field theory by a collection of `modes' or ``degrees of freedom'' $\{\phi_i\}$, where the index is both a choice among the allowed positions or momenta in the (regularized) field theory and an identification of a particular field in the theory. To compute the time evolution of an initial state at time $t$, we use a path-integral formulation to write the overlap of the time-evolved state with states of definite field value:
\begin{gather}
\braket{\tilde\phi_i}{\Psi(t)}\equiv \int^{\phi_i(\tau=t)=\tilde\phi_i}\left[D\phi_i(\tau)\right] \exp{iS\left[\phi_i(\tau)\right]} 
\end{gather}

Without changing this expression, we are free to partition the collection of modes $\{\phi_i\}$ into two non-overlapping sets, $\phi_1$ and $\phi_2$, which define a possible factorization of the (regularized) Hilbert space. We think of the first set as the system, or the modes we are interested in, and the second set as the environment, or UV modes, we do not have control over. Then we may similarly choose to do the path integral above in two steps, first integrating over one collection of degrees of freedom and then the other.
This is still a computation of overlaps between states in the full Hilbert space. However, in some circumstances we may view it as a computation of overlaps in the Hilbert space of just the fields $\phi_1$. To do so, we fix an initial state which has no initial entanglement between the two sectors, $\ket{\Psi}=\ket{\Psi_1}_1\ket{\Psi_2}_2$, where we may define the states in the 1 and 2 sectors by e.g. their overlaps with the field value states $\ket{\phi_1}$, $\ket{\phi_2}$. Then, for \emph{each} choice of initial condition $\ket{\Psi_2}$ and time-evolved state $\tilde\phi_2$, we have defined a possible time evolution of the initial state $\ket{\Psi_1}$. Typically, we have in mind the cases where the UV modes start and remain in their vacuum state (for example, if we are considering low-energy processes that should not excite heavy degrees of freedom), or perhaps have taken on some fixed external field value.

This formalism gives time-evolution rules on the Hilbert space describing the fields $\phi_1$, but in general we do not expect to relate these rules to the evolution generated by a Hamiltonian. For special, physically-relevant partitions of the degrees of freedom, and Hamiltonians on the full Hilbert space with only perturbatively weak interactions between the two sets, integration over the UV degrees of freedom does indeed give something close to unitary time evolution within the Hilbert space of $\phi_1$ alone. It is this situation that renormalization group flow describes---for example, when the division between modes is based on a momentum cutoff $\Lambda$, and the modes with momentum greater than $\Lambda$ are initialized to their vacuum state. Then, provided the time evolution has not excited the heavy modes, they only appear virtually in Feynman diagrams and may be accounted for by adding additional UV-suppressed interactions. A natural generalization, seen for example in thermal field theory, is to work with density matrices rather than states and drop the requirement that evolution neglecting the UV degrees of freedom is unitary.

In summary, RG flow, or more generally doing a portion of the path integral with specified boundary conditions, provides a (family of) maps from states in the Hilbert space of a field theory to a smaller Hilbert space describing only a subset of the original degrees of freedom. There is a sense, then, it which it might be described as coarse-graining, but it differs from the coarse-graining discussed in the body of the paper in several respects. First, unlike the channel $\mathcal{C}_\IR$ defined in Eq.\ (\ref{eq:C_IR}) above, the map described here is explicitly time-dependent. (Relatedly, the Hamiltonian on the original Hilbert space appears explicitly.) Second, the channel relies not only on the state of the full system at the final time (via the choice of $\tilde\phi_2$), but also on an initial state at some earlier time. Changing either of these choices will change the map to the coarse-grained Hilbert space, though for non-pathological actions we expect that small changes in the boundary conditions will give small changes in the nature of the map. To reiterate: the map $\mathcal{C}_\IR$ depends only on a choice of subalgebra $\mathcal{A}$, which \emph{defines} a choice of simple observables, i.e.,~what we mean by the IR degrees of freedom. But the Hamiltonian does not appear at all, except implicitly to the extent that we expect simple observables to remain simple when evolved for short times. The cost paid in the discussion here, where time evolution plays a vital role, is the loss of a unique map given a definition of the IR. It would be very interesting to attempt to incorporate time evolution more directly into the coarse-graining prescription we have discussed in this paper.

\subsection{Other comments}
The partial-trace quantum channel discussed here formalizes the intuition that the ETH keeps information about simple operators, but discards information which is not accessible to a low-energy observer. 
This is well-motivated in the existing literature, e.g.~in the study of $k$-designs as approximations to Haar-typical states, an approximation which is valid if one is only interested in data derived from lower moments of the distribution. This is something that is routinely done in for example the study of the information-theoretic aspects of a black hole, particularly in \cite{Hayden_2007}.
Another example, also in the context of black hole physics, comes from the consideration of black hole microstates. The ensemble that defines a black hole mixed state is known to obey the ETH, something which has been shown to have information-theoretic consequences \cite{Bao_2017,Bao_2019, Brand_o_2019} in terms of distinguishability and quantum error correction properties. In particular, in connection with the error correcting properties discussed in \cite{Bao_2019,Brand_o_2019}, we expect that the error correcting code maps the physical Hilbert space pre-quantum channel to the logical Hilbert space post quantum channel, and thus has strong encoding properties. However, because of the difficulty of distinguishing these states, this may also demonstrate that decoding of these states is challenging. 

Relatedly, the notion of trace distance for ETH ensembles has been studied in the context of chaotic CFTs \cite{Lashkari:2016vgj,Lashkari:2017hwq}. It would be interesting to consider whether our distinguishability arguments could be adapted to this context. In the context of holography, our algebraic setup resembles that used to discuss the notion of bulk states as an error-correcting code \cite{Almheiri:2014lwa,Harlow:2016vwg}. We hope to pursue this connection in further work.


The results of this work show that, as we might expect, the choice of kept observables in a coarse-graining quantum channel defines which ensembles of density matrices obey the ETH conditions with respect to these observables. One must consider both the ensemble and the channel to determine whether the ensemble achieves the requisite compression of the trace distance. In particular, it would be interesting in future work to consider choices of ensembles and observables that have more complicated relations to the Hamiltonian than microcanonical distributions in an energy band.

Relatedly, a drawback of our simple model is that eigenstates in different sectors have strictly vanishing overlaps for simple operators. The smoothness of the ETH envelope functions $f_\calO^{(\alpha)}$ then forces eigenstates at the edge of a sector $\calH_\alpha$ to have overlap suppressed by $e^{-S}$, so that they are zero at order $e^{-S/2}$.
This means that we cannot interpret the $\alpha$ as microcanonical bands, since this suppression does not agree with the results of a random microcanonical draw.
The simplest modification would be weak coupling between sectors, a possibility we intend to consider in future work.

Our converse result shows that exponential difficulty of distinction implies that simple operators have $O(1)$ diagonal and $O(e^{-S/2})$ off-diagonal matrix elements, as per the ETH.
However, this argument does not establish the mean or variance for these off-diagonal elements, i.e. the statistics of the $A_{ij}$ when viewed as random draws from the energy ensemble.
By considering trace distance suppression between more elaborate ensembles, it might be possible to determine these statistics from search constraints.
This suggests the tantalizing possibility that ETH distinguishability is a complete problem for some quantum complexity class, perhaps QMA, when combined with the existing hardness results, provided that the scaling is not dominated by going from pure states to mixed states.

Finally, the analysis presented here is 
reminiscent of the discussion of Petz recovery \cite{Cotler:2017erl, Chen:2019gbt, Petz:1986tvy, Junge:2015lmb, Almheiri:2017fbd, Chen:2019iro, Penington:2019kki, Almheiri:2019qdq}, where one considers the possibility of reconstructing a density matrix $\rho_{ABC}$ with access only to $\rho_{AB}$. While Petz recovery is concerned with the ability to reconstruct the full $\rho_{ABC}$, the distinguishability question asked here performs a simpler task, that of distinguishing within some discrete given set $\rho_{{ABC}_i}$. It would be interesting to consider if there is an intermediate situation where full Petz recovery is not possible, but this distinguishability task is.

\begin{acknowledgments}
We thank James Sully, Elizabeth Crosson, and Mark Van Raamsdonk for useful discussions. We would like to thank Adam Bouland for useful comments on the draft. N.B. is supported by the National Science Foundation under grant number 82248-13067-44-PHPXH, by
the Department of Energy under grant number DE-SC0019380, and by the Computational
Science Initiative at Brookhaven National Laboratory. J.P. is supported in part by the Simons Foundation and in part by
the Natural Sciences and Engineering Research Council of Canada.
D.W. is supported by an International Doctoral Fellowship from the University of British Columbia. E.W. is supported by the Berkeley Connect Fellowship.

\end{acknowledgments} 

\bibliographystyle{jhep}
\bibliography{grover-eth}

\providecommand{\href}[2]{#2}\begingroup\raggedright\begin{thebibliography}{10}

\bibitem{Srednicki1994}
M.~Srednicki, {\it Chaos and quantum thermalization},  {\em Phys. Rev. E} {\bf
  50} (Aug, 1994) 888--901.

\bibitem{Deutsch2018}
J.~M. Deutsch, {\it Eigenstate thermalization hypothesis},  {\em Reports on
  Progress in Physics} {\bf 81} (July, 2018).

\bibitem{DAlessio:2016rwt}
L.~D'Alessio, Y.~Kafri, A.~Polkovnikov, and M.~Rigol, {\it {From quantum chaos
  and eigenstate thermalization to statistical mechanics and thermodynamics}},
  {\em Adv. Phys.} {\bf 65} (2016), no.~3 239--362.

\bibitem{Deutsch1991}
J.~M. Deutsch, {\it Quantum statistical mechanics in a closed system},  {\em
  Phys. Rev. A} {\bf 43} (Feb, 1991) 2046--2049.

\bibitem{Holevo1973}
A.~S. Holevo, {\it Bounds for the quantity of information transmitted by a
  quantum communication channel},  {\em Probl. Peredachi Inf.} {\bf 9} (1973)
  3--11.

\bibitem{BBBV}
C.~H. Bennett, E.~Bernstein, G.~Brassard, and U.~Vazirani, {\it {Strengths and
  weaknesses of quantum computing}},  {\em SIAM J. Comput.} {\bf 26} (1997)
  1510--1523.

\bibitem{Rigol2012}
M.~Rigol and M.~Srednicki, {\it Alternatives to eigenstate thermalization},
  {\em Phys. Rev. Lett.} {\bf 108} (Mar, 2012) 110601.

\bibitem{Rigol2008}
M.~Rigol, V.~Dunjko, and M.~Olshanii, {\it Thermalization and its mechanism for
  generic isolated quantum systems},  {\em Nature} {\bf 452} (April, 2008)
  854--858.

\bibitem{Santos2010}
L.~F. Santos and M.~Rigol, {\it Localization and the effects of symmetries in
  the thermalization properties of one-dimensional quantum systems},  {\em
  Phys. Rev. E} {\bf 82} (Sep, 2010) 031130.

\bibitem{Santos2010b}
L.~F. Santos and M.~Rigol, {\it Onset of quantum chaos in one-dimensional
  bosonic and fermionic systems and its relation to thermalization},  {\em
  Phys. Rev. E} {\bf 81} (Mar, 2010) 036206.

\bibitem{Khatami2013}
E.~Khatami, G.~Pupillo, M.~Srednicki, and M.~Rigol, {\it
  Fluctuation-dissipation theorem in an isolated system of quantum dipolar
  bosons after a quench},  {\em Phys. Rev. Lett.} {\bf 111} (Jul, 2013) 050403.

\bibitem{Rigol2009}
M.~Rigol, {\it Breakdown of thermalization in finite one-dimensional systems},
  {\em Phys. Rev. Lett.} {\bf 103} (Sep, 2009) 100403.

\bibitem{Rigol2009b}
M.~Rigol, {\it Quantum quenches and thermalization in one-dimensional fermionic
  systems},  {\em Phys. Rev. A} {\bf 80} (Nov, 2009) 053607.

\bibitem{Steinigeweg2013}
R.~Steinigeweg, J.~Herbrych, and P.~Prelov\ifmmode~\check{s}\else \v{s}\fi{}ek,
  {\it Eigenstate thermalization within isolated spin-chain systems},  {\em
  Phys. Rev. E} {\bf 87} (Jan, 2013) 012118.

\bibitem{Beugeling2014}
W.~Beugeling, R.~Moessner, and M.~Haque, {\it Finite-size scaling of eigenstate
  thermalization},  {\em Phys. Rev. E} {\bf 89} (Apr, 2014) 042112.

\bibitem{Kim2014}
H.~Kim, T.~N. Ikeda, and D.~A. Huse, {\it Testing whether all eigenstates obey
  the eigenstate thermalization hypothesis},  {\em Phys. Rev. E} {\bf 90} (Nov,
  2014) 052105.

\bibitem{Steinigeweg2014}
R.~Steinigeweg, A.~Khodja, H.~Niemeyer, C.~Gogolin, and J.~Gemmer, {\it Pushing
  the limits of the eigenstate thermalization hypothesis towards mesoscopic
  quantum systems},  {\em Phys. Rev. Lett.} {\bf 112} (Apr, 2014) 130403.

\bibitem{Khodja2015}
A.~Khodja, R.~Steinigeweg, and J.~Gemmer, {\it Relevance of the eigenstate
  thermalization hypothesis for thermal relaxation},  {\em Phys. Rev. E} {\bf
  91} (Jan, 2015) 012120.

\bibitem{Beugeling2015}
W.~Beugeling, R.~Moessner, and M.~Haque, {\it Off-diagonal matrix elements of
  local operators in many-body quantum systems},  {\em Phys. Rev. E} {\bf 91}
  (Jan, 2015) 012144.

\bibitem{Biroli2010}
G.~Biroli, C.~Kollath, and A.~M. L\"auchli, {\it Effect of rare fluctuations on
  the thermalization of isolated quantum systems},  {\em Phys. Rev. Lett.} {\bf
  105} (Dec, 2010) 250401.

\bibitem{Roux2010}
G.~Roux, {\it Finite-size effects in global quantum quenches: Examples from
  free bosons in an harmonic trap and the one-dimensional bose-hubbard model},
  {\em Phys. Rev. A} {\bf 81} (May, 2010) 053604.

\bibitem{Sorg2014}
S.~Sorg, L.~Vidmar, L.~Pollet, and F.~Heidrich-Meisner, {\it Relaxation and
  thermalization in the one-dimensional bose-hubbard model: A case study for
  the interaction quantum quench from the atomic limit},  {\em Phys. Rev. A}
  {\bf 90} (Sep, 2014) 033606.

\bibitem{Neuenhahn2012}
C.~Neuenhahn and F.~Marquardt, {\it Thermalization of interacting fermions and
  delocalization in fock space},  {\em Phys. Rev. E} {\bf 85} (Jun, 2012)
  060101.

\bibitem{Khatami2012}
E.~Khatami, M.~Rigol, A.~Rela\~no, and A.~M. Garc\'{\i}a-Garc\'{\i}a, {\it
  Quantum quenches in disordered systems: Approach to thermal equilibrium
  without a typical relaxation time},  {\em Phys. Rev. E} {\bf 85} (May, 2012)
  050102.

\bibitem{Genway2012}
S.~Genway, A.~F. Ho, and D.~K.~K. Lee, {\it Thermalization of local observables
  in small hubbard lattices},  {\em Phys. Rev. A} {\bf 86} (Aug, 2012) 023609.

\bibitem{Mondaini2016}
R.~Mondaini, K.~R. Fratus, M.~Srednicki, and M.~Rigol, {\it Eigenstate
  thermalization in the two-dimensional transverse field ising model},  {\em
  Phys. Rev. E} {\bf 93} (Mar, 2016) 032104.

\bibitem{Dymarsky:2016ntg}
A.~Dymarsky, N.~Lashkari, and H.~Liu, {\it {Subsystem ETH}},  {\em Phys. Rev.
  E} {\bf 97} (2018) 012140, [\href{http://arxiv.org/abs/1611.08764}{{\tt
  arXiv:1611.08764}}].

\bibitem{wedderburn1964lectures}
J.~Wedderburn, {\em Lectures on Matrices}.
\newblock American mathematical society colloquium publications. Dover
  Publications, 1964.

\bibitem{Kabernik2020}
O.~Kabernik, J.~Pollack, and A.~Singh, {\it Quantum state reduction:
  Generalized bipartitions from algebras of observables},  {\em Phys. Rev. A}
  {\bf 101} (Mar, 2020) 032303.

\bibitem{Balachandran2013}
A.~P. Balachandran, T.~R. Govindarajan, A.~R. de~Queiroz, and A.~F. Reyes-Lega,
  {\it Algebraic approach to entanglement and entropy},  {\em Phys. Rev. A}
  {\bf 88} (Aug, 2013) 022301.

\bibitem{qctt}
D.~{Deutsch}, {\it {Quantum theory, the Church-Turing principle and the
  universal quantum computer}},  {\em Proceedings of the Royal Society of
  London Series A} {\bf 400} (July, 1985) 97--117.

\bibitem{jaynes1}
E.~T. Jaynes, {\it Information theory and statistical mechanics},  {\em Phys.
  Rev.} {\bf 106} (May, 1957) 620--630.

\bibitem{katz1967}
A.~Katz, {\em Principles of Statistical Mechanics: The Information Theory
  Approach}.
\newblock W. H. Freeman, 1967.

\bibitem{NandC}
M.~A. Nielsen and I.~L. Chuang, {\em Quantum Computation and Quantum
  Information: 10th Anniversary Edition}.
\newblock Cambridge University Press, USA, 10th~ed., 2011.

\bibitem{Page_1993}
D.~N. Page, {\it Average entropy of a subsystem},  {\em Physical Review
  Letters} {\bf 71} (Aug, 1993) 1291–1294.

\bibitem{reiss2007}
R.~Reiss and M.~Thomas, {\em Statistical Analysis of Extreme Values: With
  Applications to Insurance, Finance, Hydrology and Other Fields}.
\newblock Birkhaeuser Verlag Basel - Boston - Berlin, 2007.

\bibitem{Knill_2000}
E.~Knill, R.~Laflamme, and L.~Viola, {\it Theory of quantum error correction
  for general noise},  {\em Physical Review Letters} {\bf 84} (Mar, 2000)
  2525–2528.

\bibitem{Brand_o_2019}
F.~G. Brandão, E.~Crosson, M.~B. Şahinoğlu, and J.~Bowen, {\it Quantum error
  correcting codes in eigenstates of translation-invariant spin chains},  {\em
  Physical Review Letters} {\bf 123} (Sep, 2019).

\bibitem{Grover1996}
L.~K. Grover, {\it A fast quantum mechanical algorithm for database search},
  in {\em Proceedings of the Twenty-eighth Annual ACM Symposium on Theory of
  Computing}, pp.~212--219.
\newblock ACM, New York, NY, USA, 1996.

\bibitem{Bao2016}
N.~Bao, A.~Bouland, and S.~P. Jordan, {\it Grover search and the no-signaling
  principle},  {\em Phys. Rev. Lett.} {\bf 117} (Sep, 2016) 120501.

\bibitem{Biham_2002}
E.~Biham and D.~Kenigsberg, {\it Grover’s quantum search algorithm for an
  arbitrary initial mixed state},  {\em Physical Review A} {\bf 66} (Dec,
  2002).

\bibitem{Hayden_2007}
P.~Hayden and J.~Preskill, {\it Black holes as mirrors: quantum information in
  random subsystems},  {\em Journal of High Energy Physics} {\bf 2007} (Sep,
  2007) 120–120.

\bibitem{Bao_2017}
N.~Bao and H.~Ooguri, {\it Distinguishability of black hole microstates},  {\em
  Physical Review D} {\bf 96} (Sep, 2017).

\bibitem{Bao_2019}
N.~Bao and N.~Cheng, {\it Eigenstate thermalization hypothesis and approximate
  quantum error correction},  {\em Journal of High Energy Physics} {\bf 2019}
  (Aug, 2019).

\bibitem{Lashkari:2016vgj}
N.~Lashkari, A.~Dymarsky, and H.~Liu, {\it {Eigenstate Thermalization
  Hypothesis in Conformal Field Theory}},  {\em J. Stat. Mech.} {\bf 1803}
  (2018), no.~3 033101.

\bibitem{Lashkari:2017hwq}
N.~Lashkari, A.~Dymarsky, and H.~Liu, {\it {Universality of Quantum Information
  in Chaotic CFTs}},  {\em JHEP} {\bf 03} (2018) 070.

\bibitem{Almheiri:2014lwa}
A.~Almheiri, X.~Dong, and D.~Harlow, {\it {Bulk Locality and Quantum Error
  Correction in AdS/CFT}},  {\em JHEP} {\bf 04} (2015) 163,
  [\href{http://arxiv.org/abs/1411.7041}{{\tt arXiv:1411.7041}}].

\bibitem{Harlow:2016vwg}
D.~Harlow, {\it {The Ryu--Takayanagi Formula from Quantum Error Correction}},
  {\em Commun. Math. Phys.} {\bf 354} (2017), no.~3 865--912,
  [\href{http://arxiv.org/abs/1607.03901}{{\tt arXiv:1607.03901}}].

\bibitem{Cotler:2017erl}
J.~Cotler, P.~Hayden, G.~Penington, G.~Salton, B.~Swingle, and M.~Walter, {\it
  {Entanglement Wedge Reconstruction via Universal Recovery Channels}},  {\em
  Phys. Rev. X} {\bf 9} (2019), no.~3 031011.

\bibitem{Chen:2019gbt}
C.-F. Chen, G.~Penington, and G.~Salton, {\it {Entanglement Wedge
  Reconstruction using the Petz Map}},  {\em JHEP} {\bf 01} (2020) 168.

\bibitem{Petz:1986tvy}
D.~Petz, {\it {Sufficient subalgebras and the relative entropy of states of a
  von Neumann algebra}},  {\em Commun. Math. Phys.} {\bf 105} (1986), no.~1
  123--131.

\bibitem{Junge:2015lmb}
M.~Junge, R.~Renner, D.~Sutter, M.~M. Wilde, and A.~Winter, {\it {Universal
  Recovery Maps and Approximate Sufficiency of Quantum Relative Entropy}},
  {\em Annales Henri Poincare} {\bf 19} (2018), no.~10 2955--2978.

\bibitem{Almheiri:2017fbd}
A.~Almheiri, T.~Anous, and A.~Lewkowycz, {\it {Inside out: meet the operators
  inside the horizon. On bulk reconstruction behind causal horizons}},  {\em
  JHEP} {\bf 01} (2018) 028.

\bibitem{Chen:2019iro}
Y.~Chen, {\it {Pulling Out the Island with Modular Flow}},  {\em JHEP} {\bf 03}
  (2020) 033.

\bibitem{Penington:2019kki}
G.~Penington, S.~H. Shenker, D.~Stanford, and Z.~Yang, {\it {Replica wormholes
  and the black hole interior}},  \href{http://arxiv.org/abs/1911.11977}{{\tt
  arXiv:1911.11977}}.

\bibitem{Almheiri:2019qdq}
A.~Almheiri, T.~Hartman, J.~Maldacena, E.~Shaghoulian, and A.~Tajdini, {\it
  {Replica Wormholes and the Entropy of Hawking Radiation}},  {\em JHEP} {\bf
  05} (2020) 013.

\end{thebibliography}\endgroup

\end{document}